\documentclass[aps,prl,twocolumn,amsmath,amssymb]{revtex4}

\usepackage{bm}
\usepackage{graphicx}

\newcommand{\be}{\begin{eqnarray}}
\newcommand{\ee}{\end{eqnarray}}
\newcommand{\tr}{{\operatorname{Tr}}}
\newcommand{\proj}[1]{{\left|#1\rangle\!\langle #1\right|}}

\begin{document}

\title{Information Causality as a Physical Principle}

\author{Marcin Paw{\l}owski$^{1}$, Tomasz Paterek$^2$, Dagomir Kaszlikowski$^2$, Valerio Scarani$^2$, Andreas Winter$^{3,2}$ and Marek \.Zukowski$^1$}

\affiliation{
 1) Institute of Theoretical Physics and Astrophysics, University of Gda\'nsk, 80-952 Gda\'nsk, Poland
 2) Centre for Quantum Technologies and Department of Physics, National University of Singapore, 3 Science Drive 2, 117543 Singapore, Singapore
 3) Department of Mathematics, University of Bristol, Bristol BS8 1TW, United Kingdom
}

\begin{abstract}
Quantum physics exhibits remarkable distinguishing characteristics.
For example, it gives only probabilistic predictions (non-determinism) and
does not allow copying of unknown states (no-cloning\cite{woo82}).
Quantum correlations may be stronger than any classical ones\cite{BELL},
nevertheless information cannot be transmitted faster than light (no-signaling).
However, all these features do not single out quantum physics.
A broad class of theories exist which share such traits  with quantum mechanics,
while they allow even stronger than quantum correlations\cite{PR}.
Here, we introduce the principle of Information Causality.
It states that information that Bob can gain about a previously completely unknown
to him data set of Alice, by using all his local resources
(which may be correlated with her resources)
and a classical communication from her, is bounded by
the information volume of the communication.
In other words, if Alice communicates $m$ bits to Bob, the total information access that Bob gains to her data is not greater than $m$. For $m=0$, Information Causality reduces to the standard no-signaling principle.
We show that this new principle is respected both in classical and quantum physics,
whereas it is violated by all the no-signaling correlations
which are stronger that the strongest quantum correlations.
Maximally strong no-signalling correlations would
allow Bob access to {\em any} $m$ bit subset of the whole data set held by Alice.
If only one bit is sent by Alice ($m=1$), this is tantamount to Bob being
able to access the value of {\em any} single bit of Alice's data
(but of course not all of them).
We suggest that Information Causality, a generalization of no-signaling,
might be one of the foundational properties of Nature.
\end{abstract}

\maketitle

Classical (as opposed to quantum) physics rests on the assumption that all physical quantities have well defined values simultaneously. Relativity is based on clear-cut physical statements: the speed of light and the electric charge are the same for all observers. In contradistinction, the definition of quantum physics is still rather a description of its formalism: the theory in which systems are described by Hilbert spaces and dynamics is reversible. This situation is all the more unexpected as quantum physics is the most successful physical theory and quite a lot is known about it. Some of its counter-intuitive features are almost a popular knowledge: all scientists, and many laymen as well, know that quantum physics predicts only probabilities, that some physical quantities (such as position and momentum) cannot be simultaneously well defined and that the act of measurement generically modifies the state of the system. Entanglement and no-cloning are rapidly claiming their place in the list of well-known quantum features; coming next in the queue are the feats of quantum information such as the possibility of secure cryptography\cite{qkd1,qkd2} or the teleportation of unknown states\cite{teleport}.

These features are so striking, that one could hope some of them provide the physical ground behind the formalism. Is quantum physics, for instance, the most general theory that allows violations of Bell inequalities, while satisfying no-signaling? The question was first asked by Popescu and Rohrlich\cite{PR} and the answer was found to be negative: impossibility of being represented in terms of local variables is a property shared by a broad class of \textit{no-signaling theories}. Such theories predict intrinsic randomness, no-cloning\cite{mas06,bar07}, an information-disturbance trade-off\cite{sca07} and allow for secure cryptography\cite{mas08-1,mas08-2,mas08-3}. As for teleportation and entanglement swapping\cite{ENT-SWAP}, after a first negative attempt\cite{SHORT-GISIN-POPESCU}, it seems that they can actually be defined as well within the general no-signaling framework\cite{BARNUM,SKRZ-BRUNNER-POPESCU}. In summary, most of the features that have been highlighted as ``typically quantum'' are actually shared by all possible no-signaling theories. Only a few discrepancies have been noticed: some no-signaling theories would lead to an implausible simplification of distributed computational tasks\cite{VANDAM,BRASSARD,LPSW,skrz2} and would exhibit very limited dynamics\cite{barrett}.
This state of affairs highlights the importance of the no-signaling principle but leaves us still rather in the dark about the specificity of quantum theory.

In the present paper we define and study a previously unnoticed feature, which we call \textit{Information Causality}. Information Causality generalizes no-signaling and is respected by both classical and quantum physics.
However, as we shall show, it is violated by all no-signaling theories that are endowed with
correlations which are
stronger than the strongest quantum correlations. It can therefore be used as a principle to distinguish physical theories from nonphysical ones and
is a good candidate to be one of the foundational assumptions
which are at the very root of quantum theory.

Formulated as a principle, Information Causality states that \textit{information gain that Bob can reach about previously unknown to him data set of Alice, by using all his local resources and $m$ classical bits communicated by Alice, is at most $m$ bits}. The standard no-signaling condition is just Information Causality for $m=0$. It is important to keep in mind that the principle assumes classical communication: if quantum bits were allowed to be transmitted the information gain could be higher as demonstrated in the quantum super-dense coding protocol\cite{DENSE_CODING}.
The efficiency of this protocol is based on the use of quantum entanglement
and Information Causality holds true even if the quantum bits are transmitted
provided they are disentangled from the systems of the receiver.
This follows from the Holevo bound,
which limits information gain after transmission of $m$ such qubits
to $m$ classical bits.

We demonstrate that in a world in which
certain tasks are ``too simple'' (compare with Refs.\cite{VANDAM,BRASSARD}),
and there exists implausible accessibility of remote data,
Information Causality is violated.
Consider a generic situation in which Alice has a database of $N$ bits described by a string $\vec a$.
She would like to grant Bob access to as big portion of the database as possible
within fixed amount of classical communication.
If there were no pre-established correlations between them,
communication  of $m$ bits would open access to at most $m$ bits of the database.
With pre-shared correlations they could expect to do better
(but, as we shall show, in the real world they would be mistaken).
For concreteness, consider a \emph{generic} task illustrated in Figure 1. It is
%(AW: changed phrase) similar to
a distributed version of random access coding\cite{AMBAINIS-RAC,VERSTEEG-WEHNER}, oblivious transfer\cite{WOLF-WULLSCHLEGER,SHORT-GISIN-POPESCU} and related communication complexity problems\cite{BRASSARD_CCP}.
Alice receives a string of $N$ random and independent bits, $\vec a = (a_0, a_1, ..., a_{N-1})$.
Bob receives a random value of $b=0,...,N-1$,
and is asked to give a value of the $b$th bit of Alice
after receiving from her a message of $m$ classical bits.
The restrictions are only on the communication that can take place \emph{after} the inputs have been provided. The resources that Alice and Bob may have shared in advance are assumed to be no-signaling because allowing signaling resources would open other communication channels. In a classical world, these additional resources would be correlated lists of bits; in a quantum world, Alice and Bob may share an arbitrary quantum state. But the task itself is open to accommodate any hypothetical resource producing no-signaling correlations, even such that go beyond the possibilities of quantum physics. We shall call these imaginary resources \textit{no-signaling boxes}, in short NS-boxes.
%AW: added sentence:
The impact of stronger-than-quantum correlations on the efficiency of random access coding has been studied recently from a different angle\cite{VERSTEEG-WEHNER}.

\begin{figure}
\includegraphics[scale=0.5]{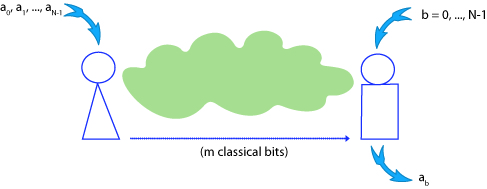}
\caption{
\textbf{The task.} Alice receives $N$ random and independent bits $\vec{a}=(a_0,a_1,...,a_{N-1})$. In a separate location, Bob receives a random variable $b\in\{0, 1,..., N-1\}$. Alice sends $m$ classical bits to Bob with the help of which Bob is asked to guess the value of the $b$th bit in the Alice's list, $a_b$. Alice and Bob can share any no-signaling resources.
  Information Causality limits the efficiency of solutions to this task.
  It bounds the mutual information between Alice's data
  and all that Bob has at hand after receiving the message.
}
\label{task}
\end{figure}

Clearly, there exists a protocol which allows Bob to give the correct value of at least $m$ bits.
If Alice sends him an $m$-bit message $\vec x = (a_0,...,a_{m-1})$ Bob will guess $a_b$ perfectly whenever $b\in\{0,...,m-1\}$. The price to pay is that he is bound to make a sheer random guess for $b\in\{m,...,N-1\}$. Since the pre-shared correlations contain no information about $\vec{a}$, for every strategy there will be tradeoff between the probabilities to guess different bits of $\vec{a}$.
Let us denote Bob's output by $\beta$. The efficiency of Alice's and Bob's strategy can be quantified by
\begin{eqnarray}
 I &\equiv& \sum_{K=0}^N I(a_K : \beta | b = K)\,
 \label{I}
\end{eqnarray}
where $I(a_K : \beta | b = K)$ is the Shannon mutual information
between $a_K$ and $\beta$, computed under the condition that Bob has
received $b=K$\cite{}.
One can also show that
\begin{equation}
  I  \geq \, N-\sum_{K=0}^N h(P_K),
 \label{XXXX}
\end{equation}
where $h(x) = - x \log_2 x - (1-x) \log_2(1-x)$ is the binary entropy of $x$,
and $P_K$ is the probability that $a_K=\beta$, again for the case of $b=K$.
To get the inequality, the $a_K$ have been assumed to be unbiased and independently
distributed (see details in the Supplementary Information).

Ideally, we would like to define that Information Causality holds if
after transferring the $m$-bit message, the mutual information
between Alice's data $\vec{a}$ and everything that Bob has, i.e.~the
message $\vec{x}$ and his part $B$ of the pre-shared correlation,
is bounded by $m$. As intuitively appealing such a definition is,
it has the severe issue that it is not theory-independent. Specifically,
a mutual information expression ``$I(\vec{a} : \vec{x}, B)$'' has to be defined
for a state involving objects from the underlying nonlocal theory
(the possibilities include classical correlation, a shared
quantum state, NS-boxes, etc.). It is far from clear whether mutual
information can be defined consistently for all nonlocal correlations,
nor whether such a definition would be unique.

Instead, we shall show that \emph{if} a mutual information can be defined
that obeys three elementary properties, then (a) Information Causality holds
and (b) $I(\vec{a}:\vec{x},B) \geq I$. Thus we obtain the following
necessary condition for Information Causality:
\begin{equation}
  I \le m.\label{ICcond}
\end{equation}
We stress that the parameter $I$ is independent of any underlying physical theory:
$I$ does not involve any details of a particular physical model
but is fully determined by Alice's and Bob's input bits and Bob's output.
In this sense it resembles Bell's parameter\cite{BELL},
which also involves only random variables and
can be used to test different physical theories.

For a system composed of parts $A$, $B$, $C$, prepared in a state
allowed by the theory, we need to assign symmetric and
non-negative mutual informations $I(A:B)$, etc. The elementary properties
mentioned above are the following.

\noindent
\textit{(1) Consistency:} If the subsystems $A$ and $B$ are both classical, then
$I(A:B)$ should coincide with Shannon's mutual information.

\noindent
\textit{(2) Data processing inequality:} Acting on one of the parts locally by any
state transformation allowed in the theory cannot increase the mutual information.
I.e., if $B \rightarrow B'$ is a permissible map between systems, then
\(
  I(A:B) \geq I(A:B').
\)
This says that any local manipulation of data can only decay information.

\noindent
\textit{(3) Chain rule:} There exists a \emph{conditional mutual information}
$I(A:B|C)$ such that the following identity is satisfied for all states and
triples of parts:
\(
  I(A:B,C) = I(A:C) + I(A:B|C).
\)
Note that this implies an identity between ordinary mutual informations:
\[
  I(A:B,C) - I(A:C) = I(A:B|C) = I(A,C:B) - I(B:C).
\]

%AW:Removed the following
%The quantity $I$ gives a lower bound on the mutual information
%between all Alice's data and data available to Bob from the message and his part of the no-signaling resource:
%$ I(\vec a : \vec x, \vec s)\ge I$,
%where $\vec x$ denotes message from Alice
%and $\vec s$ is a state of Bob's part of the resource.
%For the proof (see details in the Supplementary Information),
%we recursively apply the chain rule to obtain
%$I(\vec a : \vec{x}, \vec s) \ge \sum_{K=1}^N I(a_K : \vec{x},\vec s)$. Next, we note
%that Bob's output bit $\beta$ is obtained from the data $\vec{x}$ and $\vec s$ at
%his disposal; since information cannot increase by data processing\cite{NIELSEN-CHUANG},
%one has $I(a_K:\vec{x},\vec s)\ge I(a_K:\beta|b=K)$.
%By definition, Information Causality is fulfilled if $I(\vec a : \vec{x}, \vec s) \le m$,
%and therefore the necessary condition for Information Causality to hold becomes

Information Causality holds both in classical and quantum physics;
we may focus on the latter because the former is a special case of it. This is because
one can define quantum mutual information in formal extension of Shannon's
quantity, using von Neumann entropy\cite{CERF-ADAMI}, and all three of the
above properties are fulfilled\cite{NIELSEN-CHUANG}. Details can be found
in the Supplementary Information, but in a nutshell one argues as follows:

To show (a), denote by $B$ Bob's quantum system holding the shared quantum state $\rho_{AB}$,
Alice's data $\vec{a} = (a_0,\ldots,a_{N-1})$, and the $m$-bit message $\vec{x}$;
our objective is to prove $I(\vec{a} : \vec{x},B) \leq m$.
First, the chain rule for mutual information yields
$I(\vec a : \vec{x}, B) = I(\vec a : B) + I(\vec a : \vec{x} | B)$.
Second, $I(\vec a : B) = 0$ because without the message Alice's data and
Bob's quantum state are independent (expressing the no-signaling condition).
Third, we use chain rule again to express the conditional mutual information
as $I(\vec a : \vec{x} | B) = I(\vec{x} : \vec{a},B) - I(\vec{x}:B)
\leq I(\vec{x}:\vec{a},B)$. Finally, the latter can be upper
bounded by $I(\vec{x}:\vec{x})\leq m$, invoking data processing.
Similarly, (b) is obtained by repeated application of the chain rule,
data processing inequality and non-negativity of mutual information
(see the Supplementary Information for details).

In order to study how other no-signaling theories can violate Information Causality,
we focus on the necessary condition (\ref{ICcond}).
First consider the simplest example of two-bit input of Alice, $(a_0, a_1)$;
it is described in Figure 2.
The probability that Bob correctly gives the value of the bit $a_{0}$ is
\begin{equation}
P_\mathrm{I}=\frac{1}{2}\left[P(A\oplus B=0|0,0)+P(A\oplus B=0|1,0)\right],
\label{PI}
\end{equation}
and the analogous probability for the bit $a_{1}$ reads
\begin{equation}
P_\mathrm{II} = \frac{1}{2}\left[P(A\oplus B=0|0,1)+P(A\oplus B=1|1,1)\right],
\label{PII}
\end{equation}
where the symbol $\oplus$ denotes summation modulo $2$.

\begin{figure}
\includegraphics[scale=0.5]{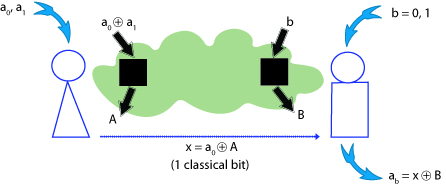}
\caption{
\textbf{van Dam's protocol\cite{VANDAM} (see also Wolf and Wullschleger\cite{WOLF-WULLSCHLEGER}).}
This is the simplest case in which Information Causality can be violated.
Alice receives two bits $(a_{0},a_{1})$ and is allowed to send only one bit to Bob.
A convenient way of thinking about no-signaling resources is to consider paired black boxes
shared between Alice and Bob (NS-boxes).
The correlations between inputs $a,b = 0,1$ and outputs $A,B = 0,1$ of the boxes are
described by probabilities $P(A \oplus B = ab | a,b)$.
The no-signaling is satisfied due to uniformly random local outputs.
With suitable NS-boxes Alice and Bob violate Information Causality.
She uses $a=a_{0}\oplus a_{1}$ as an input to the shared NS-box and obtains the outcome $A$,
which is used to compute her message bit $x=a_{0}\oplus A$ for Bob.
Bob, on his side, inputs $b=0$ if he wants to learn $a_{0}$, and $b=1$ if he wants to learn $a_{1}$; he gets the outcome $B$.
Upon receiving $x$ from Alice, Bob computes his guess $\beta=x\oplus B=a_{0}\oplus A\oplus B$.
The probability that Bob correctly gives the value of the bit $a_{0}$ is
$P_\mathrm{I}=\frac{1}{2}\left[P(A\oplus B=0|0,0)+P(A\oplus B=0|1,0)\right]$,
and the analogous probability for the bit $a_{1}$ reads
$P_\mathrm{II} = \frac{1}{2}\left[P(A\oplus B=0|0,1)+P(A\oplus B=1|1,1)\right]$,
which follow by inspection of the different cases.}
\label{figww}
\end{figure}

One can recognize that these probabilities are intimately linked with
the Clauser-Horne-Shimony-Holt parameter\cite{CHSH} $S$,
which can be used to quantify the strength of correlations.
Indeed,
\begin{equation}
  S = \sum_{a=0}^1 \sum_{b=0}^1 P(A\oplus B=ab|a,b)=2\left(P_\mathrm{I}+P_\mathrm{II}\right).
\end{equation}
The classical correlations are bounded by $S\leq S_{C}=3$ (the equivalent form of Bell inequality\cite{BELL,CHSH}).
Quantum correlations exceed this limit up to $S\leq S_{Q}= 2+\sqrt{2}$ (the so-called Tsirelson bound\cite{TSIRELSON}).
The maximal algebraic value of $S_{NS}=4$ is reached by the Popescu-Rohrlich (PR) box\cite{PR}, which is an extremal no-signaling resource. PR-boxes maximally violate Information Causality because they predict $P_\mathrm{I}=P_\mathrm{II}=1$, i.e. $I=2$ for $m=1$, so here occurs an
extreme violation of Information Causality.
Bob can learn perfectly \textit{either} bit.
$I = 2$ measures the sum-total of the information accessible to Bob.
However, he cannot learn \textit{both} Alice's bits -- the latter would imply signaling.

The protocol works just as well for any Boolean function of the inputs, $f(\vec a, b)$.
It is sufficient that Alice inserts to her PR-box the sum of $f(\vec a,0) \oplus f(\vec a, 1)$.
If Information Causality is maximally violated, Bob can learn the value of $f(\vec a,b)$
for any one of his inputs, irrespectively of Alice's input data.
Even more surprisingly, this is so also if he does not know the function to be computed.

We shall now demonstrate that Information Causality is violated as soon as the quantum Tsirelson limit for the CHSH inequality is exceeded. This result of ours can be also seen as an information-theoretic proof of the Tsirelson bound, independent of the formalism of Hilbert spaces, relying instead only
on the existence of a consistent information calculus for certain correlations.

First we note that, using a suitable local randomization procedure that does not change the value of the parameter $S$, any NS-box can be brought to a simple form\cite{mas06}: the local outcomes are uniformly random and the correlations are given by
\begin{eqnarray}
  P(A\oplus B=ab|a,b)&=&\frac{1}{2}(1+E),
  \label{isotropic}
\end{eqnarray}
with $0\leq E\leq 1$. The case $E=1$ corresponds to the PR-box; $E=0$ describes uncorrelated random bits. The classical bound $S\le S_C$ is violated as soon as $E>\frac{1}{2}$; the Tsirelson bound of quantum physics becomes $E\leq E_Q=\frac{1}{\sqrt{2}}$, attained by performing suitable measurements on the singlet state of two two-level systems\cite{BELL,TSIRELSON}.

\begin{figure}
\includegraphics[scale=0.25]{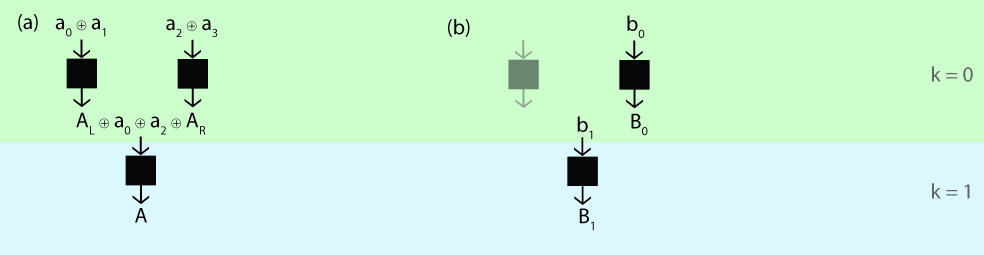}
  \caption{
  \textbf{Information Causality identifies the strongest quantum correlations}.
  The possible no-signaling correlations satisfying Information Causality
  can be precisely identified using the depicted scheme.
  Alice receives $N = 2^n$ input bits
  and correspondingly Bob receives $n$ input bits $b_n$ which describe the index of the bit he is interested in, $b = \sum_{k=0}^{n-1} b_k 2^{k}$.
  She is allowed to send a single bit, $m=1$.
  In the case of $n=2$, to encode information about her data,
  Alice uses a pyramid of NS-boxes as shown in the panel (a).
  Note that Fig. 2 shows how Bob can correctly
  guess the first or second bit of Alice using a single pair of the boxes (the case of $n=1$).
  If Alice has more bits, then they recursively use this protocol in the following way.
  E. g., for four input bits of Alice, two pairs of NS-boxes on the level $k=0$ allow Bob to make the guess of a value
  of any one of Alice's bits as soon as he knows either $a_0 \oplus A_L$ or $a_2 \oplus A_R$,
  where $A_L$ ($A_R$) is the output of her left (right) box on the level $k=0$,
  which are the one-bit messages of the protocol in Fig. 2.
  These can be encoded using the third box, on the level $k=1$, by inserting their sum to the Alice's box
  and sending $x = a_0 \oplus A_L \oplus A$ to Bob ($A$ is the output of her box on the level $k=1$).
  Depending on the bit he is interested in, he now reads a suitable message using the box on the level $k=1$
  and uses one of the boxes on the level $k=0$.
  An example of situation in which Bob aims at the value of $a_2$ or $a_3$ is depicted in the panel (b).
  Bob's final answer is $x \oplus B_0 \oplus B_1$, where $B_k$ is the output of his box on the $k$th level.
  Generally, Alice and Bob use a pyramid of $N-1$ pairs of boxes placed on $n$ levels.
  Looking at the binary decomposition of $b$ Bob aims ($n-r$) times at the left bit and $r$ times at the right, where $r = b_0 + ... + b_{n-1}$.
  His final guess is the sum of $\beta = x \oplus B_0 \oplus ... \oplus B_{n-1}$.
  Therefore, Bob's final guess is correct whenever he has made an even number of errors in the intermediate steps.
  This leads to Eq. (\ref{PK}) for the probability of his correct final guess (see Supplementary Information for the details of this calculation).
}
\label{figwwtree}
\end{figure}

The bound that Information Causality imposes on correlations can be identified
using a pyramid of NS-boxes and nesting the simple protocol described above (see Figure 3). Now Alice receives $N=2^n$ bits and the probability that Bob guesses $a_K$ correctly is given by
\begin{equation}
P_K = \frac{1}{2}\left[1+E^{n}\right].
\label{PK}
\end{equation}
Inserting this expression into (\ref{I}), one finds that the Information Causality condition $I\leq 1$ is violated as soon as $2E^2>1$ and $n$ large enough, i.e. $E>E_Q$.
Since all NS-boxes can be brought to the form (\ref{isotropic}) without changing the value of $S$, we conclude indeed that every NS-box with stronger than quantum correlations violates the Information Causality condition.
In Supplementary Information the more general result is proved, that for any
$\frac{1}{2}\left(E_\mathrm{I}^2+E_\mathrm{II}^2\right) > E_Q^2$ where $E_j=2P_j-1$ -- see
eqs. (\ref{PI}) and (\ref{PII}) -- Information Causality is violated, and conversely if it
is fulfilled, that there exists a quantum correlation with these probabilities.

In conclusion, we have identified the principle of Information Causality, which precisely distinguishes physically realized correlations from nonphysical ones (in the sense that quantum mechanics cannot reach them). It is phrased in operational terms and in a theory-independent way and therefore we suggest it is at the same foundational level as the no-signaling condition itself, of which it is a generalization.

The new principle is respected by all correlations accessible with quantum physics while it excludes all no-signaling correlations, which violate the quantum Tsirelson bound.
Among the correlations that do not violate that bound it is not known whether Information Causality singles out exactly those allowed by quantum physics.
If it does, the new principle would acquire even stronger status.

We thank Matthias Christandl, Vlatko Vedral and Stephanie Wehner for stimulating discussions.
This work was supported by the National Research Foundation and the Ministry of Education in Singapore,
and by the European Commission through IP ``QAP''.
AW acknowledges support by the U.K. EPSRC through the ``QIP IRC'' and an Advanced Fellowship, by a Royal Society Wolfson Merit Award, and a Philip Leverhulme Prize.

\section{SUPPLEMENTARY INFORMATION}

\section{The generic nature of the considered task }

Assume that Alice has a data set, a CD or whatever, which
can be encoded into a $N$-bit string, $\vec{a}$.
Bob may wish to have an access to that set of data.
Of course, without any communication he has no access at all.
However, if they share randomness, or a source of randomness, and a protocol, Alice can, by transferring $m$ bits, allow him an access to a specific $m$-bit subsequence of her data. Thus $N-m$ bits are still inaccessible to Bob. Transfer of $m$-bits reduces the number of inaccessible bits from $N$ to $N-m$, or more if the protocol is not optimal. We have an accessibility gain of up to $m$ bits.
PR-boxes clearly violate this limit. A transfer of $m$-bits, due to no-signaling still allows to access $m$ bit sequences only, however the full data set is open for such a readout. The number of inaccessible bits is reduced to $0$. Accessibility gain is $N$.

In the most elementary case of just one bit transfer, the Information Causality allows one in an optimal protocol to decode the value of just one specific bit. For PR-boxes one bit transfer opens access to {\em any} bit. That is, all bits are readable, with no-signaling constraining the actual readout to just one of them.

Further, note that,
any Boolean function of Alice's and Bob's data can be put
in the following way
$$f(\vec{a},b)=\sum_{b'}\delta_{bb'}f'_{\vec{a}}({b'}),$$
where $f'_{\vec{a}}({b'})$ is again Boolean. For example, the original data string is a simple function of this kind $A_{\vec{a}}({b})=a_b$.
Therefore any function $f'_{\vec{a}}(b)$ is just a preparation of a new string of data, in form of a Boolean function of the old string. It has exactly the same length. Thus our problem contains within itself a completely general  problem of obtaining the value of {\em any} $f(\vec{a},b)$. That is, it is a generic problem for dichotomic functions.

\section{Information calculus and Information Causality}

Here we prove bounds for the mutual information
$I(\vec a : \vec x, B)$, where $\vec a$ is a
string of Alice's bits, $\vec x$ is her classical message
and $B$ denotes Bob's part of the pre-shared no-signalling resource.
assuming the three abstract properties for the mutual information,
as follows.

\noindent
\textit{(1) Consistency:} If the subsystems $A$ and $B$ are both classical, then
$I(A:B)$ should coincide with Shannon's mutual information.

\noindent
\textit{(2) Data processing inequality:} Acting on one of the parts locally by any
state transformation allowed in the theory cannot increase the mutual information.
I.e., if $B \rightarrow B'$ is a permissible map between systems, then
\(
  I(A:B) \geq I(A:B').
\)
This says that any local manipulation of data can only decay information.

\noindent
\textit{(3) Chain rule:} There exists a \emph{conditional mutual information}
$I(A:B|C)$ such that the following identity is satisfied for all states and
triples of parts:
\(
  I(A:B,C) = I(A:C) + I(A:B|C).
\)
Note that this implies an identity between ordinary mutual informations:
\[
  I(A:B,C) - I(A:C) = I(A:B|C) = I(A,C:B) - I(B:C).
\]

We show first
$I(\vec a : \vec x, B) \geq I = \sum_{K=0}^{N-1} I(a_K : \beta)$
in the case of independent Alice's input bits. Namely, by the
chain rule (3), we can isolate Alice's first bit, obtaining
$I(a_0,\ldots,a_{N-1} : \vec{x}, B) = I(a_0 : \vec{x}, B) + I(a_1,\ldots,a_{N-1} : \vec{x},B | a_0)$.
The second term on the right-hand side equals, using chain rule once
more,
$I(a_1,\ldots,a_{N-1} : \vec{x},B|a_0) = I(a_1,\ldots,a_{N-1} : \vec{x}, B, a_0) - I(a_1,\ldots,a_{N-1}:a_0)$,
in which, due to the independence of Alice's inputs, $I(a_1,\ldots,a_{N-1}:a_0)=0$. Applying
the data processing inequality (2) to the first term here then implies
\begin{equation}
  I(\vec a : \vec{x}, B) \ge I(a_0 : \vec{x}, B) + I(a_1,\ldots,a_{N-1} : \vec{x},B).
\end{equation}
Iterating these steps $N-1$ times to the rightmost information gives
\begin{equation}
  \label{IND}
  I(\vec a : \vec{x}, B) \ge \sum_{K=0}^{N-1} I(a_K : \vec{x}, B).
\end{equation}
Finally, we observe that Bob's guess bit $\beta$ is obtained at the end
from $b$, $\vec{x}$ and $B$. Hence,
the data processing inequality puts
a limit of $I(a_K:\beta|b=K) \leq I(a_K:\vec{x},B)$ on Bob's
accessible information. Putting this together with eq.~(\ref{IND})
yields the result,
\begin{equation}
  I(\vec a : \vec{x}, B) \ge   \sum_{K=0}^{N-1} I(a_K:\beta | b= K) \equiv I,
  \label{INDEP}
\end{equation}
which is the efficiency described in the main text. Note that implicitly we have
made use of the consistency condition (1) here.

\medskip
Second, we prove that the same assumptions lead to
$I(\vec{a} : \vec{x},B) \leq m$, i.e.~Information Causality.
To do so we need a little preparation.
Note that from consistency (1) and data processing (2),
we inherit automatically an important property of Shannon
mutual information, namely the fact that $I(A:B) = 0$
if two systems $A$ and $B$ are independent, i.e.~the state is
a (tensor) product of states on $A$ and on $B$, respectively.
To prove this, observe that the state can thus be
prepared by allowed local operations starting from
two \emph{classical} independent systems. These have
zero mutual information by consistency (1), so $I(A:B) \leq 0$
by data processing (2). On the other hand, mutual information
must be non-negative, hence $I(A:B) = 0$.

Now,
\[\begin{split}
  I(\vec{a} : \vec{x},B) &= I(\vec{a}:B) + I(\vec{a} : \vec{x}|B) \\
                         &= I(\vec{a} : \vec{x}|B) \\
                         &= I(\vec{x} : \vec{a},B) - I(\vec{x} : B) \\
                         &\leq I(\vec{x} : \vec{a},B),
\end{split}\]
where we have invoked the chain rule (3), the independence of $\vec{a}$
from $B$ (which owes itself to the no-signaling condition), chain
rule once more and non-negativity of mutual information.

We are finished now once we argue that $I(\vec{x} : \vec{a},B) \leq I(\vec{x}:\vec{x})$,
because the latter is a quantity only involving classical objects, so it
can be evaluated as the Shannon entropy of $\vec{x}$ by the consistency requirement (1),
and the entropy is upper bounded by $m$. But this inequality
follows once more from data processing (2), because the joint
state of $\vec{x}$, $\vec{a}$ and $B$ is given by some distribution
on $\vec{x}$, and a joint state for $\vec{a}$ and $B$ for each
value $\vec{x}$ can take. In other words, there is a state
preparation for each value of $\vec{x}$, hence there must exist the
corresponding state transformation $\vec{x} \rightarrow \vec{a},B$
in the theory.

\section{Simplified lower bound on $I$}

The conditional mutual informations can be simplified using the probability of Bob's correct guess
of $a_K$, denoted by $P_K$, i.e. the probability that $a_K \oplus \beta = 0$, given $b=K$.
Since Alice's inputs are uniformly random, the binary entropy $h(a_K) = 1$,
we have $I(a_K:\beta | b= K) = 1 - H(a_K | \beta, b = K)$.
Note that, $H(a_K | \beta, b = K) = H(a_K \oplus \beta| \beta, b = K)$
because knowing $\beta$ leaves the same uncertainty about $a_K$ and $a_K \oplus \beta$
(this can also be proved using the chain rule for conditional entropy).
Omitting the conditioning on $\beta$ can only increase the entropy,
$H(a_K \oplus \beta| \beta, b = K) \leq  H(a_K \oplus \beta| b = K) = h(P_K)$.
Therefore,
\begin{equation}
I \ge N - \sum_{K=0}^{N-1} h(P_K),
\label{NPK}
\end{equation}
as stated in the main text.
This inequality can also be seen as a special case of Fano's inequality \cite{NIELSEN-CHUANG}.
In a more general case, in which Alice's inputs acquire values from an alphabet of $d$ elements,
Fano's inequality gives the bound
\begin{equation}
I \ge N \log_2 d - \sum_{K=0}^{N-1} h(P_K) - \sum_{K=0}^{N-1} (1-P_K) \log_2 (d-1).
\end{equation}
Similarly, one can write a bound for any inputs of Alice.

Since the necessary condition for Information Causality to hold reads $I \le m$,
one finds by looking at the expression (\ref{NPK}) that Information Causality limits
the probability of Bob's correct guess, unless all information about Alice's bits
is communicated to Bob:
\begin{equation}
\sum_{K=0}^N h(P_K) \ge N-m.
\end{equation}

\section{Information causality in classical and quantum physics}
\label{appa}

Here we show that Information Causality holds
in classical and quantum physics. All we have to do, in
the light of our previous reasoning, is to write down expressions
for the mutual information and conditional mutual information,
and confirm that they satisfy properties (1)--(3).
We focus on quantum correlations
because classical correlations form a subset of quantum correlations.
With respect to any tripartite state $\rho_{ABC}$, denote by
$\rho_A$, etc. its reduced states, and write $S(\rho) = -\tr\rho\log\rho$
for the von Neumann entropy. Then let
\begin{align*}
  I(A:B)   &= S(\rho_A)+S(\rho_B)-S(\rho_{AB}), \\
  I(A:B|C) &= S(\rho_{AC})+S(\rho_{BC})-S(\rho_{ABC})-S(\rho_C).
\end{align*}
Both expressions are manifestly invariant under swapping $A$
and $B$, and non-negative by strong subadditivity\cite{NIELSEN-CHUANG}.
Clearly, consistency (1) holds, as classical correlations
are embedded as matrices diagonal in some fixed local bases
and then von Neumann entropy reduces to Shannon entropy.
Also the chain rule (3) is an easily verified identity.
The data processing inequality (2) is equivalent once more
to the data processing inequality\cite{NIELSEN-CHUANG}.

\medskip
To verify the steps in our abstract derivation of Information
Causality in the quantum case, denote the initial state shared
between Alice and Bob by $\rho_{AB}$.
Including Alice's data as orthogonal states of a reference
system $R$, the situation before the communication can be
described by the state
\be
  \frac{1}{2^N} \sum_{\vec{a}\in\{0,1\}^N} \proj{\vec{a}}_R \otimes \rho_{AB}.
\ee
For each value of $\vec{a}$
Alice has to perform local operations to obtain the message $\vec{x}$
she wants to send to Bob. Whatever her algorithm to do so, it can
be condensed into a quantum measurement (POVM)
$(M_{\vec{x}}^{(\vec{a})})_{\vec{x}\in\{0,1\}^m}$, and so the
joint state of Alice's data, the message (represented by orthogonal states of
a ``message'' system $X$) and Bob's system is given by
\be
  \frac{1}{2^N} \sum_{\vec{a}\in\{0,1\}^N} \proj{\vec{a}}_R \otimes
        \sum_{\vec{x}\in\{0,1\}^m} \tr_A\bigl( \rho_{AB}(M_{\vec{x}}^{(\vec{a})}\otimes\openone \bigr).
\ee

\section{Tsirelson bound from information causality}
\label{appb}

We present a proof that
Information Causality is violated by all stronger than quantum correlations.
Our protocol of the main proof uses NS-boxes,
which produce uniformly random local outcomes
and correlations described by the probabilities
$P(A \oplus B = a b | a,b)$,
where $a,b = 0,1$ are the inputs to the boxes and $A,B = 0,1$
are the outputs for Alice and Bob respectively.
It will be sufficient to consider the situation where
Alice communicates only one bit to Bob, i.e. $m=1$.
The number of Alice's input bits is chosen as $N = 2^n$,
where $n$ is an integer parameterizing the task.
Correspondingly, Bob receives $n$ input bits to encode the index
$b$ as a binary string $(b_0, b_1,\ldots, b_{n-1})$,
i.e. $b = \sum_{k=0}^{n-1} b_k 2^{k}$.

We generalize the procedure given in the main text to $N$ bits, recursively,
using the insight of~\cite{VANDAM} that any function, which can
be written as a Boolean formula with ANDs,
XORs and NOTs, can be computed in a distributed manner
using the same number of PR-boxes \cite{PR} as ANDs and one bit of communication.
The function we are considering is $f_n(\vec{a},b) \equiv a_b$, with
$\vec{a} = (a_0, a_1, \ldots a_{N-1})$.
In the simplest case, $n=1$, the function of the task reads
\begin{equation}
  f_1\bigl( (a_0,a_1), b_0 \bigr) = a_0 \oplus b_0(a_0 \oplus a_1).
\end{equation}
It involves a single AND. Alice inputs $a_0 \oplus a_1$ into the PR-box,
Bob $b_0$; with her output $A$ Alice forms the message $x = a_0 \oplus A$,
so that Bob can obtain $x \oplus B = a_b$.

Moving to $n > 1$, write $\vec{a} = \vec{a}'\vec{a}''$ with two
bit-strings $\vec{a}'$ and $\vec{a}''$ of length $N/2 = 2^{n-1}$ each.
Then it is a straightforward exercise to verify that
\begin{equation}
  f_n( \vec{a},b ) = f_{n-1}(\vec{a}',b') \oplus
                        b_{n-1} \bigl[ f_{n-1}(\vec{a}',b') \oplus f_{n-1}(\vec{a}'',b') \bigr],
  \label{f_j}
\end{equation}
where $b'$ is the string of $n-1$ Bob's bits $(b_0,...,b_{n-2})$.
Thus, if $f_{n-1}$ could be written using $N/2-1$ ANDs, this formula
expresses $f_n$ using $N-1$ AND operations. For instance,
\be \nonumber
  f_2\bigl( (a_0,a_1,a_2,a_3), (b_0, b_1) \bigr)
     = a_0 \oplus b_0(a_0 \oplus a_1)\oplus \\
              \oplus b_1 \bigl[ a_0 \oplus b_0(a_0 \oplus a_1) \oplus a_2 \oplus b_0(a_2 \oplus a_3) \bigr].
\ee
To convert this to a distributed protocol, Alice and Bob use three PR-boxes.
To the first one Alice inputs $a_0 \oplus a_1$,
to the second one $a_2 \oplus a_3$,
and to the third one $a_0 \oplus A_1 \oplus a_2 \oplus A_2$,
where $A_1$ and $A_2$ are her outputs from the first and second box, respectively.
She transmits $x = a_0 \oplus A_1 \oplus A_3$,
where $A_3$ is her output from the third box.
Depending on his inputs, Bob will use two different boxes to decode $a_b$.
He inserts the bit $b_1$ distinguishing groups $(a_0,a_1)$ and $(a_2,a_3)$
to the third box, obtaining output $B_3$.
Due to the correlations of the boxes, the sum $x \oplus B_3$
gives the value of $a_0 \oplus A_1$ or $a_2 \oplus A_2$,
depending on $b_1=0$ or $b_1=1$.
These are exactly the messages he would obtain from Alice
in the scenario with the single pair of boxes,
and the protocol is now reduced to the previous one.
For example, if $b_1=1$ he will input $b_0$ to the second PR-box,
giving him an output $B_2$, so that he can form
$x \oplus B_3 \oplus B_2 = a_2 \oplus A_2 \oplus B_2$,
which is either $a_2$ or $a_3$. Note that the other PR-box is ignored
by Bob, or he may as well input $b_0$ to it, too -- it is not
important because he doesn't need its output.

In the general case, Alice and Bob share $N-1 =  2^n-1$ PR-boxes,
and for every set of inputs Bob uses $n$ of them. The protocol
can be explained recursively, based on eq.~(\ref{f_j}): Alice
and Bob use the protocol for $n-1$ Bob's bits on the input pairs
$(\vec{a}',b')$ and $(\vec{a}'',b')$, involving $N/2-1$ PR-boxes in
each one, resulting in two single-bit messages $x'$ and $x''$
that Alice would send to Bob if their objective were to
compute $f_{n-1}$. Instead, she inputs $x' \oplus x''$ into the
last PR-box, while Bob inputs $b_{n-1}$; they obtain an output bits
$A$ and $B$, respectively, so when Alice finally sends the
message $x = x' \oplus A$, this allows Bob to obtain $x' \oplus b_{n-1}(x'\oplus x'')$,
which is either $x'$ or $x''$ depending on $b_{n-1}=0$ or $b_{n-1}=1$.
Then, the protocol for  $n-1$ bits tells him
the outputs of which PR-boxes
he should use in order to arrive at $a_b$.
Since in the protocol for $n-1$ Bob's bits
he only needs the outputs of $n-1$ boxes, he reads $n$ outputs
for $n$ bits; likewise, Alice uses $2(N/2-1)+1 = N-1$ PR-boxes
in total.

Now, we simply substitute PR-boxes with NS-boxes,
having their probabilities of guessing first and second bit of Alice's input sum given by $P_{\mathrm{I}}$ and $P_{\mathrm{II}}$, respectively.
By looking at his input bits, Bob finds that the final guess of the value of $a_b$
involves aiming $(n-k)$ times at the left bit and $k$ times at the right,
where $k = b_0 + \dots + b_{n-1}$ is the number of 1s in the binary decomposition of $b$.
Since Bob's answer is computed
as the sum of the message $x$ and suitable outputs of $n$ boxes,
whenever even number of boxes produce
``wrong'' outputs, i.e. such that $A \oplus B \ne ab$,
Bob still arrives at the correct final answer.
Therefore, Bob's guess is correct whenever he has made an even number of errors in the intermediate steps.

Let us denote by
\begin{equation}
Q_{\mathrm{even}}^{(k)}(P) = \sum_{j=0}^{\lfloor \frac{k}{2}\rfloor} {k \choose 2 j} P^{k-2j} (1-P)^{2j} = \frac{1}{2}[1 + (2P-1)^k]
\end{equation}
the probability to make an even number of errors
when using $k$ pairs of boxes, each producing a correct value with probability $P$.
Similarly, the probability to make an odd number of errors reads
\be \nonumber
Q_{\mathrm{odd}}^{(k)}(P)= \sum_{j=0}^{\lfloor \frac{k-1}{2}\rfloor} {k \choose 2 j+1} P^{k-2j-1} (1-P)^{2j+1} 
\\
=\frac{1}{2}[1-(2P-1)^k ].
\ee
With this notation, the probability that Bob's final guess of the value of $a_K$ is correct is given by
\begin{eqnarray}
P_K&=&Q_{\mathrm{even}}^{(n-k)}(P_{\mathrm{I}})\, Q_{\mathrm{even}}^{(k)}(P_{\mathrm{II}})+Q_{\mathrm{odd}}^{(n-k)}(P_{\mathrm{I}})\, Q_{\mathrm{odd}}^{(k)}(P_{\mathrm{II}})\nonumber\\ &=& \frac{1}{2}\left[1+E_\mathrm{I}^{n-k} E_\mathrm{II}^{k}\right] \label{PK-evenodd}
\end{eqnarray}
with $E_j=2P_j-1$.

We are ready to compute the Information Causality quantity (1) of the main text:
\begin{eqnarray}
  I &=& \sum_{K=1}^N [1-h(P_K)]\nonumber\\&=&\sum_{k=0}^n {n \choose k} \left[1-h\left(\frac{1+E_\mathrm{I}^{n-k} E_\mathrm{II}^{k}}{2}\right)\right]\nonumber\\
  &\ge & \frac{1}{2\ln 2}\,\sum_{k=0}^n {n \choose k}(E_\mathrm{I}^2)^{n-k} (E_\mathrm{II}^2)^{k}\nonumber\\ &=& \frac{1}{2\ln 2}\,\left(E_\mathrm{I}^2+E_\mathrm{II}^2\right)^n
\end{eqnarray}
where we have used $1-h\left(\frac{1+y}{2}\right)\ge\frac{y^2}{2\ln 2}$. Therefore, if
\begin{eqnarray}
E_\mathrm{I}^2+E_\mathrm{II}^2&>&1,
\label{violic}
\end{eqnarray}
there exist $n$ such that $I>1$.

It is also possible, and not difficult, to show that whenever $E_\mathrm{I}$ and $E_\mathrm{II}$ do not violate Information Causality then there exists a quantum protocol that gives such correlations.

For the isotropic correlations (6) of the main text, $E_\mathrm{I}=E_\mathrm{II}=E$, hence eq. (\ref{PK-evenodd}) becomes $P_K=\frac{1}{2}\left[1+E^{n}\right]$ and eq. (\ref{violic})
becomes $2E^2>1$ as stated in the main text.

\end{document}